\documentclass[prl,
preprint,
preprintnumbers,
showpacs
]{revtex4}
\usepackage {graphicx}
\usepackage{bm} % bold math

\begin{document}

\title{Thermally Enhanced Neutralization in Hyperthermal Energy Ion Scattering}

\author{C.E. Sosolik}
\email[Corresponding author:  ]{sosolik@nist.gov}
\altaffiliation[Present Address:  ]{National Institute of Standards and 
Technology, Gaithersburg, MD, 20899-8412}
\author{J.R. Hampton}
\author{A.C. Lavery}
\altaffiliation[Present Address:  ]{Woods Hole Oceanographic Institution, 
Woods Hole, MA 02543}
\author{B.H. Cooper}
\altaffiliation[ ]{Deceased.}
\affiliation{
Laboratory of Atomic and Solid State Physics, Cornell
University, Ithaca, New York 14853-2501}
\author{J.B. Marston}
\affiliation{Department of Physics, Box 1843, Brown University, Providence,
Rhode Island 02912}
\date{\today}
\begin{abstract}
Neutralization probabilities are presented for hyperthermal energy
Na$^+$ ions scattered from a Cu(001) crystal as a function of surface
temperature and scattered velocity.  A large enhancement in
neutralization is observed as the temperature is increased.
Velocity-dependent charge transfer regimes are probed by varying the
incident energy, with the most prominent surface temperature effects
occurring at the lowest energies.  The data agree well with results
obtained from a model based on the Newns-Anderson Hamiltonian, where
the effects of both temperature and velocity are incorporated. 
\end{abstract}
\maketitle

Measurements of the charge states of alkali ions scattered from metal
surfaces at hyperthermal energies ($\approx$ 1 eV to 1 keV) have
achieved remarkable success in isolating factors that govern
%control the magnitude and rate of 
neutralization at
surfaces \cite{Cooper:BlueBook}.  From these studies, it is well known
that the magnitude of the neutralization probability, $P_{\rm 0}$,
depends strongly on the value of the projectile ionization
potential, $I_{\rm 0}$, and the surface work function, $\Phi$, while
the rate of charge transfer is determined by the projectile-surface
coupling \cite{Note:2}.  The effects of surface temperature or $T_{\rm
S}$ on neutralization at a surface, however, have typically been
ignored experimentally. In fact, few quantitative measurements
\cite{Overbosch:1980,Bu:1990} have explored the role of $T_{\rm S}$ in
determining $P_{\rm 0}$, despite the many theoretical studies 
\cite{Shao:multiple,Merino:1998,Rasser:1980,Brako:multiple,Liu:1985,Battaglia:1986,Easa:1987,Nakanishi:1988,Sulston:multiple}
that have been devoted to the subject.  

In this Letter, we present quantitative results which show that
$T_{\rm S}$ has significant effects on the neutralization probability.
Our measurements involve scattering Na$^+$ ions from a Cu(001) surface as a
function of incident energy and $T_{\rm S}$.  The Na-Cu(001)
scattering system is interesting because both \emph{energy-dominated}
and \emph{coupling-dominated} charge transfer regimes are accessible
and can be probed by varying the incident energy or scattered
velocity\cite{Kimmel:1993b}.    In the energy-dominated regime, the
change in $P_{\rm 0}$ with scattered velocity is governed primarily by
the relative values of $I_{\rm 0}$ and $\Phi$.  In the
coupling-dominated regime, the tunneling of electrons between the
projectile and surface or equivalently, the projectile-surface
coupling, governs the velocity dependence \cite{Kimmel:1993b}.  Our
results show that increasing $T_{\rm S}$ can change $P_{\rm 0}$ in
both of these regimes,
enhancing it by as much as a factor of three at the lowest incident
energies.  The effects of $T_{\rm S}$ and scattered velocity on
$P_{\rm 0}$ have been incorporated into a quantum mechanical model
that treats the electrons involved in the charge transfer as
independent particles.  Results obtained from this model compare well
with the trends seen in our experimental data.   

Our measurements were performed in an ultra high vacuum chamber and
beamline that are described in detail elsewhere
\cite{McEachern:1988,Adler:1988a}.  All of the ion beams
were produced in a Colutron ion source that has been modified to
allow highly efficient alkali ion beam production from a solid state
source \cite{Peale:1989}.  The beams were scattered 
along the $\langle 100 \rangle$ azimuth of a Cu(001) single crystal
sample.  Surface cleanliness and long range order were monitored using
Auger electron spectroscopy and low energy electron diffraction,
respectively.  The sample temperature was varied between 200 K and
1100 K during scattering using a combination of cooling from a liquid
nitrogen reservoir connected by copper braids and heating from an
electron-beam heater mounted behind the sample.   

\begin{figure}
\includegraphics[width=7.5cm]{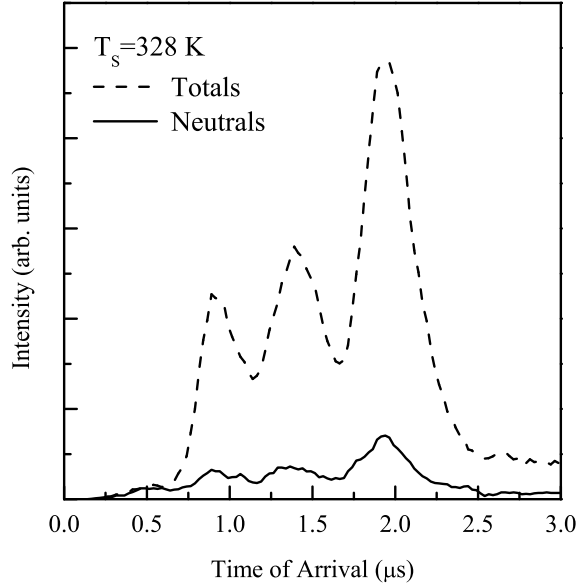}
\caption{\label{Naspectra}
Time-of-arrival spectra obtained with the NPD for 250 eV Na$^+$ ions 
scattered from  a Cu(001) crystal.  The surface temperature was held
at $328 \pm 5$ K during this measurement.  The three peaks present in
these spectra are due to trajectory types that involve scattering from
one or more individual atoms at the surface \cite{Cooper:BlueBook}.   
}
\end{figure}

Velocity-resolved charge state fractions were obtained for particles
scattered from the sample using time-of-flight techniques along with a
neutral particle detector (NPD) \cite{Kimmel:1993a}.  All measurements
were made at incident and final angles of 45$^\circ$ in the plane
defined by the incident beam direction and the surface normal.  A typical
time-of-arrival spectrum obtained using the NPD is shown in
Fig. \ref{Naspectra} for 250 eV Na$^+$ 
scattered from the Cu(001) crystal at a temperature of 328 K.  The
dashed line shows the signal due to both the neutral atoms and ions,
the total flux, and the solid line shows the signal coming from the
neutrals only.  $P_{\rm 0}$ was obtained by integrating the intensity
of the total and neutral spectra and taking a ratio of the results
\cite{Note:1}.

\begin{figure}
\includegraphics[width=7.5cm]{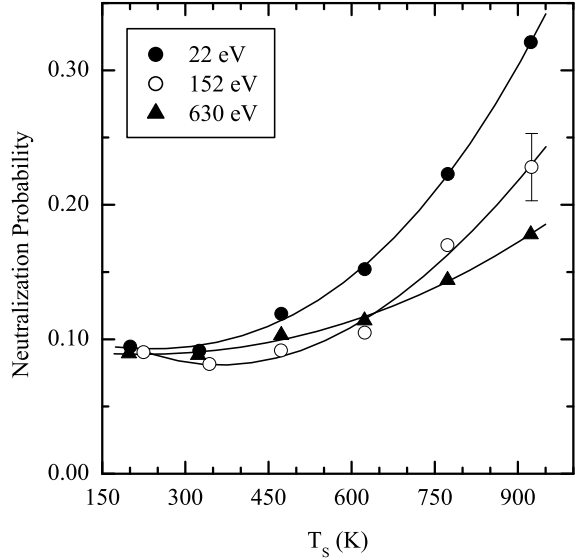}
\caption{\label{P0vsTlinear}
Temperature-dependent neutralization results for 22 eV, 152 eV, and 630 eV
Na$^+$ ions scattered from a Cu(001) crystal.  The lines are drawn to
guide the eye, and a typical error bar is shown. 
}
\end{figure}

To illustrate the dependence of $P_{\rm 0}$ on $T_{\rm S}$, typical
results obtained for incident Na$^+$ projectiles at energies of 22 eV,
152 eV, and 640 eV are shown in Fig. \ref{P0vsTlinear}.  There is a
significant increase in $P_{\rm 0}$ at all incident energies as
$T_{\rm S}$ is increased.  At each incident energy, the change in
neutralization probability is monotonic with $T_{\rm S}$, with the
largest change occurring at the lowest incident energy.  

\begin{figure}
\includegraphics[width=7.5cm]{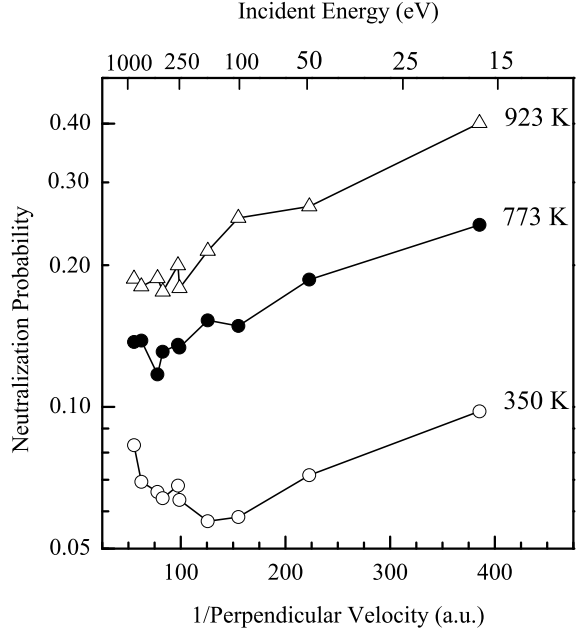}
\caption{\label{P0vsTlog}
The neutralization probability, $P_{\rm 0}$, for Na$^+$ ions scattered
from a Cu(001) crystal at surface temperatures of 350 K, 773 K, and
923 K.  $P_{\rm 0}$ is plotted on a logarithmic scale versus the 
inverse scattered perpendicular velocity to show the exponential
dependence present at low energies and high temperatures.  The
velocity is shown in atomic units (a.u.), where 1 a.u. is
approximately $2.2 \times 10^8$ cm/sec.  The corresponding incident
energy scale is noted along the top axis.   
}
\end{figure}

The dependence of $P_{\rm 0}$ on incident energy can also be observed
from the data presented in Fig. \ref{P0vsTlinear}.
Above 750 K, $P_{\rm 0}$ decreases monotonically as the incident
energy is increased.  However, below 750 K, the incident energy
dependence is non-monotonic.  This can be seen more clearly in
Fig. \ref{P0vsTlog}, where $P_{\rm 0}$ has been plotted as a function
of the inverse perpendicular velocity of the scattered projectiles at  
three different $T_{\rm S}$ values.  The non-monotonic change in
neutralization is very apparent at 350 K, the lowest temperature
shown.  
%%%%%FIGURE1%%%%%%%%%%%%%%%%%%%%%%%%%%
%%%%%%%%%%%%%%%%%%%%%%%%%%%%%%%%%%%%%%%%

To understand the velocity-~and~temperature-dependent results shown in
Figs. \ref{P0vsTlinear} and \ref{P0vsTlog} we first focus on the
energy and time dependence of charge transfer for this system at a
fixed temperature.  We show that the data collected at 350
K exhibit aspects of both energy- and coupling-dominated charge
transfer.  Then we examine the effects of increased temperature and
discuss why the effects are so large at the lowest incident energies.
Finally, we compare our measurements to the results of a theoretical
model that incorporates the effects of velocity and temperature.   

\begin{figure}
\includegraphics[width=8.0cm]{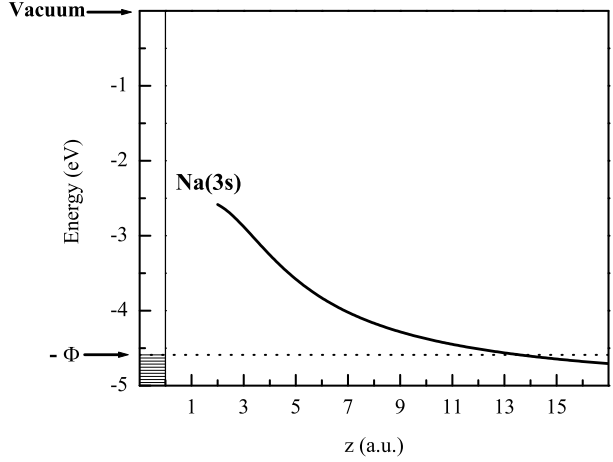}
\caption{\label{LevelDiagram}
An ionization level diagram for Na outside of a Cu(001) surface.  The
Fermi level lies at an energy $\Phi$ below the vacuum level, where
$\Phi=4.59$ eV for Cu(001).  Filled electron levels in the metal are
indicated by the lines drawn on the far left.  The distance, $z$, is
plotted in atomic units (a.u.), where 1 a.u. is 0.529 \AA.   
}
\end{figure}

The energetics of neutralization in ion-surface scattering experiments
are typically described using an ionization level diagram.  The
ionization level, $I(z)=I_{\rm 0}-1/4z$, represents both the bare
ionization potential of the projectile and the distance dependence 
induced by the presence of an image charge in the metal \cite{Note:4}.
A level 
diagram for Na ($I_{\rm 0}=5.14$ eV) outside of a Cu(001) surface is
shown in Fig. \ref{LevelDiagram}.  At any distance $z$ the
energetically-favored charge state of the Na is given by the position
of $I(z)$ relative to the Cu(001) Fermi level, $E_{\rm F}$.  At large
distances $(z>13~{\rm a.u.})$ the ionization level is below $E_{\rm F}$,
and the energetically-favored charge state is neutral.  Closer to the
surface, the positive ion is the energetically-favored charge state, as the
ionization level is shifted above $E_{\rm F}$.  
The level diagram illustrates
that different charge states are energetically favored at different
$z$ values.  This description is incomplete, however, as the
projectile-surface coupling must also be considered.
%%%%%FIGURE2%%%%%%%%%%%%%%%%%%%%%%%%%%
%%%%%%%%%%%%%%%%%%%%%%%%%%%%%%%%%%%%%%%%

The projectile-surface coupling introduces dynamical and charge state
mixing effects that play an important role in determining the
final charge state measured in a neutralization experiment.  
The dynamical effect refers to the competition that arises between the
rate of electron tunneling and the finite velocity of the scattered
projectile.  Put simply, as a projectile leaves the surface,
electrons tunnel between the projectile and 
surface, tracking the energetically favored charge state described
above.  However, the 
exponential decay of the coupling with distance implies that at some
distance $z$ the charge state is ``frozen in'' and the projectile is
left as either neutral or positive.  Quantum-mechanical mixing of
projectile charge states outside the surface, however, make this
picture more complex as they imply that there is a non-zero probability of
obtaining any 
allowed charge state at a distance $z$.  

For the Na-Cu(001) system, only the two charge states, neutral and
positive, need to be considered.  The probability of obtaining either 
charge state at a distance $z$ is determined primarily by the
energetically favored charge state at that distance.  Looking at
Fig. \ref{LevelDiagram}, we see that this implies that the positive
ion will dominate at most $z$ values.  The velocity dependence at any
$z$, however, is determined, to lowest order, by the  
relative magnitude of the projectile-surface coupling and the absolute 
energy difference between $I(z)$ and $E_{\rm F}$.
Regions where either the coupling or this energy difference are largest
determine the energy-~and~coupling-dominated charge transfer regimes,
respectively.  If a projectile leaves the surface slowly, the  
energetically-favored charge state can be tracked to large distances,
where the coupling is very small.  In this case, the velocity
dependence is classified as energy-dominated, and an exponential
dependence of $P_{\rm 0}$ on the inverse perpendicular velocity is
obtained \cite{Kimmel:1993b}.  Conversely, higher velocity projectiles
only track the energetically-favored charge state near   
the surface where the coupling is large.  This is the
coupling-dominated regime where the simple exponential dependence of
$P_{\rm 0}$ on inverse perpendicular velocity is no longer valid.  
%%%%%FIGURE3%%%%%%%%%%%%%%%%%%%%%%%%%%
%%%%%%%%%%%%%%%%%%%%%%%%%%%%%%%%%%%%%%%%

Focusing on the data taken at 350 K in Fig. \ref{P0vsTlog}, one can
see evidence for both charge transfer regimes as a function of the
inverse perpendicular velocity.  Energy-dominated charge transfer
occurs at the lowest velocities, giving $P_{\rm 0}$ an exponential
dependence for incident energies less than 100 eV.  Above 100 eV, the
exponential dependence is no longer present, and $P_{\rm 0}$
increases.  The strong projectile-surface
interaction dominates at these higher velocities, giving an
increased neutral occupancy, or as we measure it, a larger $P_{\rm 0}$
value \cite{Kimmel:1993b}.  Therefore, the non-monotonic change
observed for $P_{\rm 0}$ indicates that both energy- and 
coupling-dominated charge transfer occur in this system at 350 K.    
%%%%%FIGURE4%%%%%%%%%%%%%%%%%%%%%%%%%%
%%%%%%%%%%%%%%%%%%%%%%%%%%%%%%%%%%%%%%%%

The data in Fig. \ref{P0vsTlog} also show that the magnitude of
$P_{\rm 0}$ and its dependence on inverse perpendicular velocity
change dramatically with $T_{\rm S}$.  Generally, an increase in
$T_{\rm S}$ should alter the electron occupancy in the metal,
populating metal levels above $E_{\rm F}$ according to the Fermi-Dirac
distribution.  If $|{I(z)-E_{\rm F}}| \lesssim k_{\rm B}T_{\rm S}$, thermal
effects will become significant and lead to a larger $P_{\rm 0}$
value.  Looking at Fig. \ref{LevelDiagram}, 
we see that this condition is met at large distances, where $I(z)$
crosses $E_{\rm F}$.  This is the energy-dominated regime that is
probed at low velocities, and it explains why the measured $P_{\rm 0}$
values are very sensitive to $T_{\rm S}$ as the scattered
perpendicular velocity is decreased.  At higher velocities, $I(z)$ is
well above $E_{\rm F}$, the charge transfer is coupling-dominated, and
consequently, the effects of $T_{\rm S}$ are reduced.

We have compared our measured temperature-dependent $P_{\rm 0}$
data to results obtained with an independent particle
calculation\cite{Onufriev:1996}.  This calculation is based on the
spinless one-level Newns-Anderson 
Hamiltonian and allows for a determination of the occupancy of the Na
atomic level after scattering from the Cu surface.  It is a quantum
mechanical treatment that correctly models both of the charge transfer
regimes that are probed in our results.  This is important because
semi-classical treatments, such as the rate equations, do not
reproduce the phenomena of coupling-dominated charge transfer that
lead to the non-monotonic $P_{\rm 0}$ dependence observed in 
Fig. \ref{P0vsTlog}.  A modified Fermi-Dirac distribution
that incorporates both thermal and
velocity-smearing effects was included in the calculation.  The
distribution was obtained by performing an angular average of a
velocity-shifted Fermi-Dirac distribution over a spherical Fermi surface
and is written as
\begin{equation}
f^*(\epsilon) = {{\ln(1 + {\rm e}^{-\beta (\epsilon - e)})- \ln(1 +
{\rm e}^{-\beta (\epsilon + e)})}\over{2 \beta e}}~, 
\end{equation}
where $\beta=1/k_{\rm B}T_{\rm S}$ and $\epsilon$ is the energy of a metallic
level relative to $E_{\rm F}$.  The term $e$ is equal to $k_{\rm
F}v_{\rm proj}$, where 
$k_{\rm F}$ and $v_{\rm proj}$ are the Fermi wavevector and the
projectile velocity, respectively.  
The projectile-surface coupling used in the calculation was based on a
fit to the theoretical results of Nordlander and Tully
\cite{Nordlander:1989}. 

\begin{figure}
\includegraphics[width=7.5cm]{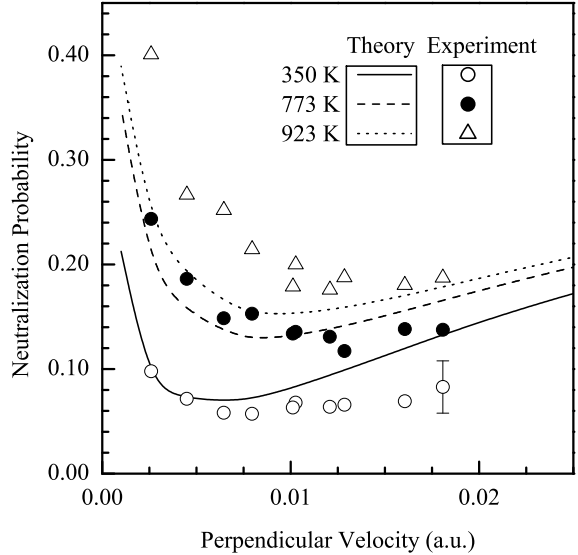}
\caption{\label{P0vsTtheory}
Comparison of the experimental neutralization probability(symbols) to the
results of the independent-particle calculation(lines) for Na$^+$ scattering
from Cu(001).  A typical error bar is shown.
}
\end{figure}

A comparison of our theoretical and experimental results for $P_{\rm
0}$ as a function of the perpendicular velocity at three different
$T_{\rm S}$ values is shown in Fig. \ref{P0vsTtheory}.  The 
calculation reproduces the increase observed in $P_{\rm 0}$
with $T_{\rm S}$ as well as the non-monotonic change seen in $P_{\rm 
0}$ with perpendicular velocity.  Also, the large increase in $P_{\rm
0}$ observed at the lowest velocity values is reproduced.
The quantitative differences evident between the experimental and
calculation results shown in this figure vary with
velocity and are very sensitive to the coupling and the inclusion of
parallel velocity effects \cite{Sosolik:Thesis}.  Although a more
complex neutralization model may be required to correctly deal with
this interplay between $T_{\rm S}$, parallel velocity, and the  
projectile-surface coupling, it is remarkable that such a simple
calculation can reproduce the observed trends.
%%%%%FIGURE5%%%%%%%%%%%%%%%%%%%%%%%%%%
%%%%%%%%%%%%%%%%%%%%%%%%%%%%%%%%%%%%%%%%

In this Letter, we have shown that $T_{\rm S}$ can have
significant effects on the $P_{\rm 0}$ values measured in hyperthermal
energy ion-surface scattering experiments.  By using Na-Cu(001) as the
experimental system, our results reveal that these thermal effects are
present for both energy- and coupling-dominated charge transfer.  The
sensitivity of these alkali ion measurements to $T_{\rm S}$ in
the energy-dominated regime demonstrate that the failure to include
$T_{\rm S}$ in the analysis or modeling of neutralization experiments
performed at finite temperatures could lead to large errors.
Furthermore, in light of these results, future experiments aimed at
measuring thermal effects due to many-body or Kondo effects for
alkaline-earth projectiles \cite{Shao:multiple,Merino:1998}, must be
carefully interpreted.    

This work was supported by the National Science Foundation
(NSF-DMR-9722771 and NSF-DMR-9712391). J.R.H. acknowledges support
from a National Physical Science Consortium Fellowship.
%%%%%%%%%%%%%%%REFERENCES%%%%%%%%%%%%%%%%%%%%%%%%%%%%%%%%%%%%%%%%


\begin{thebibliography}{23}

\bibitem{Cooper:BlueBook}
B.H. Cooper and E.R. Behringer, in \emph{Low Energy Ion-Surface 
Interactions}, edited by J.W. Rabalais (J. Wiley \& Sons, New York, 1994).

\bibitem{Note:2}
Here we refer to ``charge transfer'' as meaning the \emph{resonant}
transfer of charge between projectile and metal levels at the
same energy.  

\bibitem{Overbosch:1980} 
E.G. Overbosch, B. Rasser, A.D. Tenner, and J. Los, Surf. Sci. {\bf 92}, 
310(1980).

\bibitem{Bu:1990} 
Y. Bu, E.F. Greene, and D.K. Stewart, J. Chem. Phys. {\bf 92}, 3899(1990).

\bibitem{Shao:multiple}
H. Shao, P. Nordlander, and D.C. Langreth,
Phys. Rev. B {\bf 52}, 2988(1995);
Phys. Rev. Lett. {\bf 77}, 948(1996).

\bibitem{Merino:1998} 
J. Merino and J.B. Marston, Phys. Rev. B {\bf 58}, 6982(1998).

\bibitem{Rasser:1980} 
B. Rasser and M. Remy, Surf. Sci. {\bf 93}, 223(1980).

\bibitem{Brako:multiple}
R. Brako and D.M. Newns, Surf. Sci. {\bf 108}, 253(1981); Vacuum {\bf
32}, 39(1982). 

\bibitem{Liu:1985} 
K.C. Liu, T.F. George, and K.-S. Lam, Solid State Commun. {\bf 53}, 67(1985).

\bibitem{Battaglia:1986} 
F. Battaglia, K.C. Liu, and T.F. George, Int. J. Quantum
Chem. Suppl. {\bf 19}, 477(1986). 

\bibitem{Easa:1987} 
S.I. Easa and A. Modinos, Surf. Sci. {\bf 183}, 531(1987).

\bibitem{Nakanishi:1988} 
H. Nakanishi, H. Kasai, and A. Okiji, Surf. Sci. {\bf 197}, 515(1988).

\bibitem{Sulston:multiple}
K.W. Sulston and F.O. Goodman, J. Chem. Phys. {\bf 112}, 2486(2000);
F.O. Goodman and K.W. Sulston, J. Chem. Phys. {\bf 114}, 3265(2001).

\bibitem{Kimmel:1993b}
G.A. Kimmel and B.H. Cooper, Phys. Rev. B {\bf 48}, 12164(1993).

\bibitem{McEachern:1988}
R.L. McEachern {\it et al.}, Rev. Sci. Instrum. {\bf 59}, 2560 (1988).

\bibitem{Adler:1988a}
D.L. Adler and B.H. Cooper, Rev. Sci. Instrum. {\bf 59}, 137 (1988).

\bibitem{Peale:1989}
D.R. Peale, D.L. Adler, B.R. Litt, and B.H. Cooper, Rev. Sci. Instrum. 
{\bf 60}, 730 (1989).

\bibitem{Kimmel:1993a} 
G.A. Kimmel and B.H. Cooper, Rev. Sci. Instrum. {\bf 64}, 672 (1993).

\bibitem{Note:1}
All three peaks were used to determine $P_{\rm 0}$ as little variation
was observed for the different trajectory types represented.
The assigned velocity value was obtained by averaging over the
velocities present in each spectrum. 

\bibitem{Note:4}
$I(z)$, $I_{\rm 0}$, and $z$ are expressed here in atomic units.

\bibitem{Onufriev:1996}
A.V. Onufriev and J.B. Marston, Phys. Rev. B {\bf 53}, 13340(1996).

\bibitem{Nordlander:1989}
P. Nordlander and J.C. Tully, Surf. Sci. {\bf 211/212}, 207(1989).

\bibitem{Sosolik:Thesis}
C.E. Sosolik, Ph.D. thesis, Cornell University, 2001.

\end{thebibliography}
\end{document}